\documentclass[10pt,journal,twoside]{IEEEtran} 
\usepackage{cite}
\usepackage{amsmath,amssymb,amsfonts}
\usepackage{graphicx}
\usepackage{textcomp}
\usepackage{bm}
\usepackage{algorithm}
\usepackage{algpseudocode}

\def\BibTeX{{\rm B\kern-.05em{\sc i\kern-.025em b}\kern-.08em
		T\kern-.1667em\lower.7ex\hbox{E}\kern-.125emX}}

\title{Low Latency Stand Alone Compute-Efficient Forecasting of Marine Engine Time Series Data}

\author{Y.~Harsha~Vardhana~Reddy and Soumyendu~Raha
    \thanks{Y. Harsha Vardhana Reddy is a Post Graduate student at the Department of Computational and Data Sciences, Indian Institute of Science, Bangalore 560012, India (e-mail: Harshavyr@iisc.ac.in).}
    \thanks{Soumyendu Raha is with the Department of Computational and Data Sciences, Indian Institute of Science, Bangalore 560012, India (e-mail: raha@iisc.ac.in).}
}

\begin{document}

\maketitle
\begin{abstract}
The operational reliability of a high performance marine vessel depends critically on the health of its marine propulsion systems, which are increasingly subjected to diverse operational loads and environmental stressors. This paper proposes a robust mathematical framework for non-linear state-space forecasting of marine engine parameters using adaptive-window multi-particle stochastic differential equations. Traditional time-series models such as Vector Autoregressive Integrated Moving Average, often fail to capture the inherent stochasticity and transient dynamics of complex systems due to their reliance on fixed-window linear assumptions. To address this, we develop a dual-layered estimation approach: first, an adaptive lookback mechanism dynamically adjusts the learning window size based on the instantaneous drift magnitude, ensuring responsiveness during non-stationary regimes. Second, a Multi-Particle ensemble is evolved via Euler-Maruyama discretization, where each particle trajectory represents a stochastic realization of the system state. To refine the ensemble mean and mitigate the ``noise-chasing" behavior of raw estimators, a Girsanov transform induced change of probability measure is implemented, assigning higher probabilistic weights to particles that align with the physical drift. Theoretical evaluation and empirical benchmarking demonstrate that the proposed adaptive SDE framework significantly outperforms classical statistical baselines in multi-step prediction stability and computational efficiency. The model provides a scalable, ``grey-box" solution for real-time risk quantification in systems characterized by high-frequency volatility and non-linear transitions
\end{abstract}


\section{Introduction}
\label{intro} The operational reliability of marine  vessels are increasingly dependent on the technological maturity of its propulsion systems.To address the problem of enhanced operational envelope, the focus has shifted toward developing robust, local mathematical and machine learning models to optimize operational efficiency and cost-effectiveness. 

Modern marine vessels are powered by diverse propulsion systems, including Gas Turbines, Marine diesel engines used for both propulsion and auxiliary power generation. However, prolonged service in harsh maritime environments causes these engines to deviate from their nominal design parameters due to varying loads, environmental conditions, and mechanical wear. Traditional physics-based thermodynamic models often fail to account for these real-world deviations, necessitating a shift toward Maintenance 4.0—a paradigm that leverages data-driven tools for proactive maintenance and real-time risk quantification.

While deep learning and regression-based approaches have gained popularity, they often lack the interpretability  required for critical marine engineering decisions and can be prone to overfitting. This research proposes a hybrid framework grounded in Stochastic Differential Equations (SDEs) to capture both the deterministic physical trends of the engine and the inherent stochastically (randomness) of operational sensor data. By treating the engine states as single/multi particle moving through a state-space, the model provides significant forecast into the future states.

\section{Background}
In marine propulsion, the primary energy conversion processes are governed by the Brayton cycle for gas turbines and the Diesel cycle for internal combustion engines. While these ideal cycles provide the ``nominal" performance baseline, real-world operation introduces deviations that this research tries to address.Below is the fundamental physics behind the various engines employed onboard marine Ships.
\subsection{Gas Turbine}
The Gas turbine engine operates on the open-loop Brayton cycle. It consists of four fundamental stages that define the engine's deterministic drift:

Isentropic Compression (1-2): Ambient air is drawn into the compressor, increasing its pressure and temperature.

Constant-Pressure Heat Addition (2-3): Fuel is injected and burned in the combustion chamber.

Isentropic Expansion (3-4): High-temperature gases expand through the turbine, generating work for propulsion and powering the compressor.

Heat Rejection (4-1): Exhaust gases are released into the atmosphere

\begin{figure}[t]
	\centering
	\includegraphics[width=0.8\linewidth]{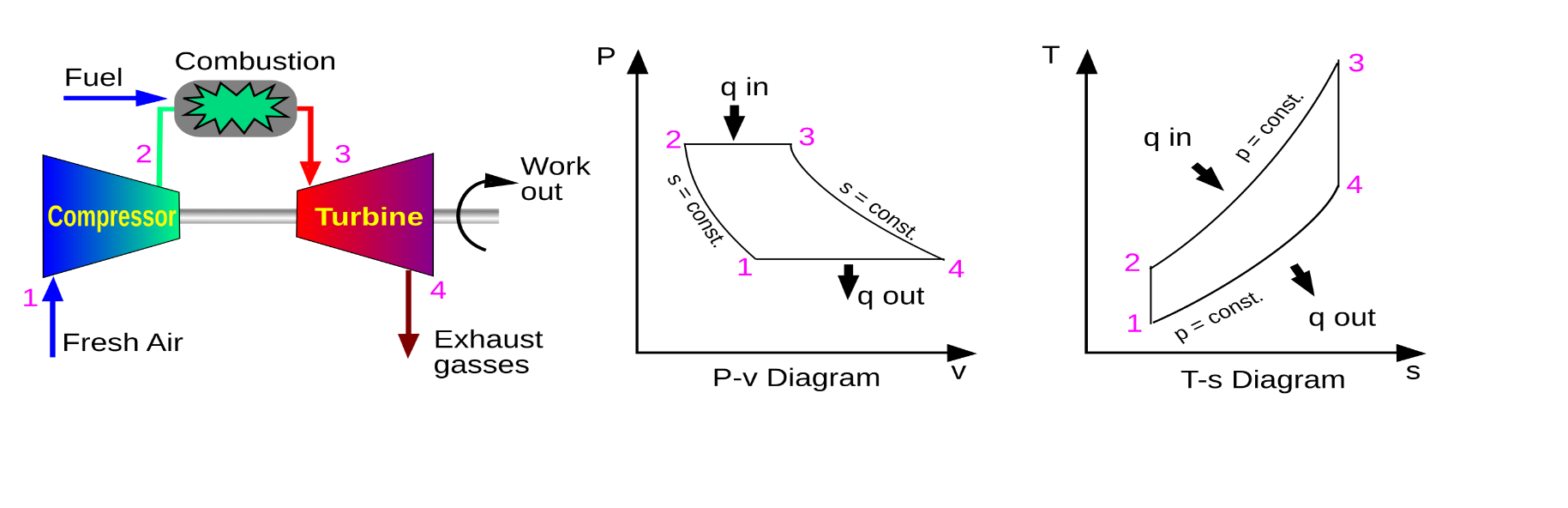}
	\caption{Gas Turbine-Brayton Cycle}
	\label{fig:Gas Turbine}
\end{figure}

\subsection{Diesel Engine}
For power generation and secondary propulsion, marine vessels rely on marine diesel engines. These operate on the four-stroke compression-ignition cycle:

Induction \& Compression (1-2): Air is compressed to a high temperature, exceeding the ignition point of the fuel.

Constant-Pressure Combustion (2-3): Fuel is injected and burns as the piston moves, maintaining nearly constant pressure.

Expansion (3-4): The power stroke where work is extracted.

Exhaust (4-1): Waste heat is expelled from the cylinder
\begin{figure}[t]
	\centering
	\includegraphics[width=0.8\linewidth]{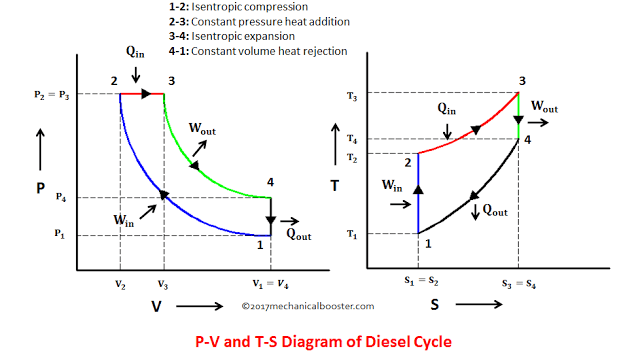}
	\caption{Diesel Cycle}
	\label{fig:Diesle Cycle}
\end{figure}
\subsection{Physics-Based vs. Data-Driven Modeling}
The fundamental challenge in marine engineering is that these ideal cycles assume ``nominal" conditions.

Physics-Based Models: These utilize the equations derived from the cycles above. While they offer high interpretability, they are often too rigid to account for ``stochastic" noise or environmental fluctuations.

Data-Driven Models: These rely on empirical sensor data. They excel at capturing trends and lack the underlying physical constraints of the Brayton or Diesel cycles.
\begin{figure}[t]
	\centering
	\includegraphics[width=0.8\linewidth]{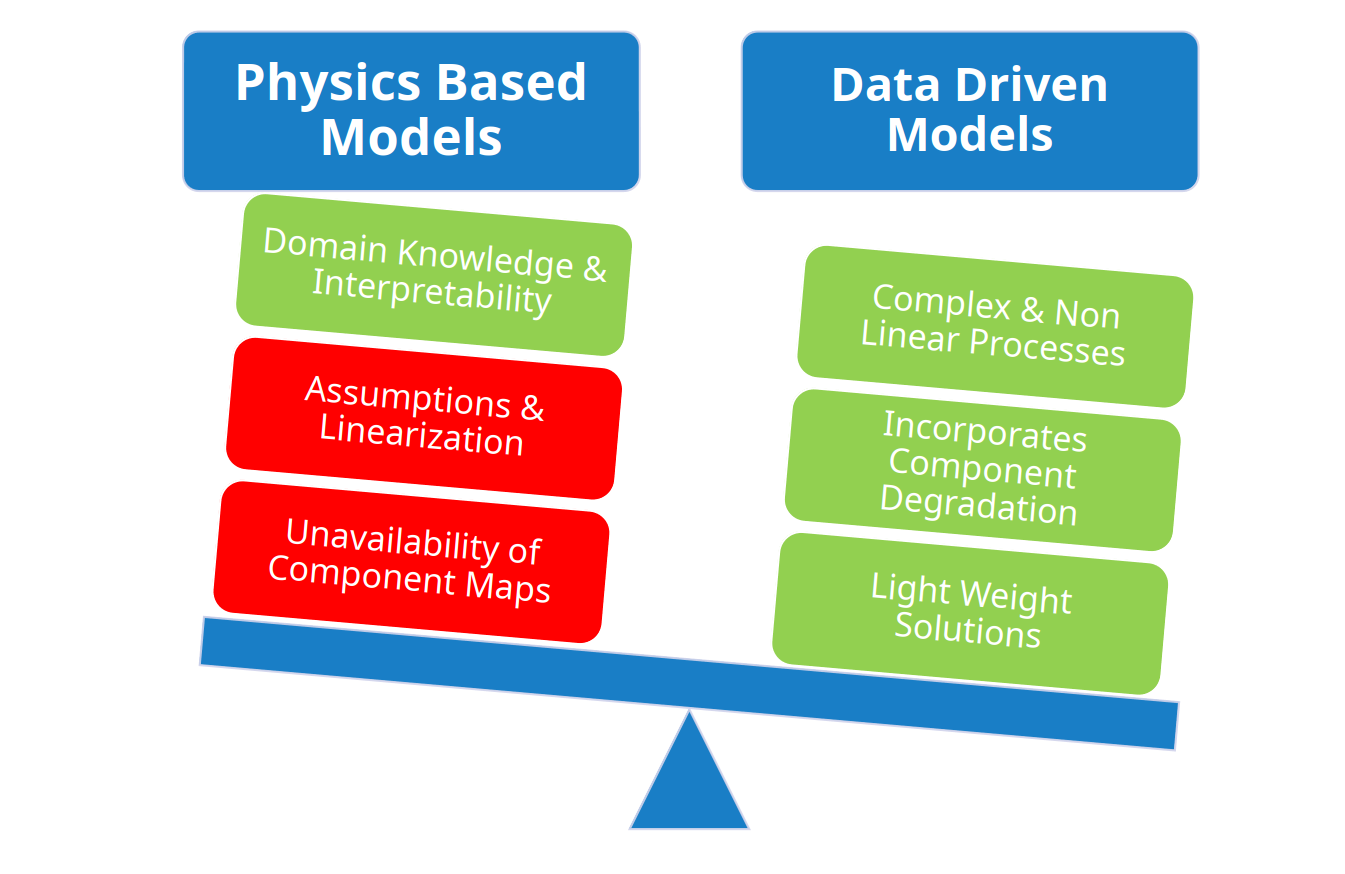}
	\caption{Data Driven Models Vs Physics Based Models}
	\label{fig:Data vs Physics}
\end{figure}
\section{Related Work}
The development of predictive frameworks for marine propulsion systems has evolved from traditional thermodynamic cycles toward high-fidelity, hybrid stochastic models.Numerous studies have explored Gas Turbine modeling and simulation, predominantly in aeronautical applications.However, limited research focuses on marine propulsion gas turbines/Diesel engines, which are crucial for marine operations.Existing approaches include physics-based models, data-driven models and hybrid models each with its advantages and limitations.
\subsection{Physics-Based and Thermodynamic Modeling}Classical performance assessment of marine engines relies on the fundamental thermodynamic cycles of Brayton and Diesel.

\cite{saravanamuttoo2007gas} Establishes the deterministic baseline for parameters such as pressure ratios and Turbine Entry Temperature (TET). While generic simulation tools allow for detailed component-level modeling, these ``White Box" approaches often struggle to account for real-world deviations caused by component aging and environmental noise.

\cite{panov2010gasturbolib} Introduced GasTurboLib, a specialized Simulink library that enables modular modeling of Gas Turbine transients.

\cite{gazzetta2017real} Developed real-time models for performance simulation, emphasizing the need for high-speed execution in diagnostic frameworks.

\cite{alexiou2005development} Utilized generic simulation tools to develop performance models that analyze Gas Turbine behavior under varying operational points.

\cite{singh2018modeling} Validated the modeling and simulation of mini Gas Turbine engines, providing a framework for smaller-scale propulsion analysis.

\subsection{Data-Driven and Machine Learning Approaches}
To address the limitations of rigid physics-based models, research has increasingly turned to ``Black Box" data-driven techniques. Regression methods and Recurrent Neural Networks (RNNs) have been successfully deployed to predict Gas Turbine parameters and fuel oil consumption.

\cite{batayev2018prediction} Demonstrated the efficacy of machine learning regression methods in predicting Gas Turbine parameters from limited sensor sets.

\cite{gkerekos2019machine} Performed a comparative study on ship main engine fuel consumption, highlighting that while machine learning excels at pattern recognition, it may fail to account for the underlying thermodynamics of the engine cycle.

\cite{asgari2021recurrent} Utilized Recurrent Neural Networks to simulate the behavior of single-shaft gas turbines, focusing on transient state accuracy.

\cite{dai2024novel} Proposed a novel data-driven approach specifically for predicting the performance degradation of gas turbines over time.

\subsection{Hybrid Modeling and Stochastic Differential Equations}

The contemporary trend in Maintenance 4.0 is the synthesis of physical knowledge with machine learning, creating ``Grey Box" models. \cite{belov2020hybrid} demonstrated that hybrid models provide superior prescriptive analytics for gas turbines by combining thermodynamic principles with empirical data. The use of Stochastic Differential Equations (SDEs) represents a sophisticated iteration of this hybrid approach.

\cite{zhigarev2021development} Developed mathematical models for diesel-generator units that incorporate stochastic variations in load, proving that SDE-based frameworks are robust for maritime power generation.

\cite{willard2023integrating} Established the ``Physics-Infused" paradigm as a means to integrate scientific knowledge with machine learning for engineering systems.

\cite{singh2023hybrid} Explored hybrid physics-infused frameworks specifically for fault diagnostics and prognostics in diesel engine cyber-physical systems.

\cite{mohammad2021hybrid} Applied hybrid physical and machine learning-oriented modeling to predict emissions in diesel compression ignition engines.

\subsection{Issues with Engineering Systems Time Series Modeling}
Overall, the following issues and insights emerge when classical thermodynamics, statistical forecasting, and modern hybrid modeling are combined and taken into consideration
\subsubsection{The Rigidity of ``White Box" Thermodynamic Models}
Research confirms that while physics-based models (Brayton and Diesel cycles) are highly interpretable, they are inherently deterministic. These models assume ``nominal" conditions and fail to account for the stochastic deviations caused by component aging, fuel quality variations, and environmental noise prevalent in maritime operations. Consequently, a purely physics-based approach lacks the agility required for real-time fault detection in degraded engines.

\subsubsection{The Interpretability Gap in ``Black Box" Machine Learning}
While machine learning models, such as those explored above, demonstrate high predictive accuracy for gas turbine/diesel engine parameters, they suffer from a lack of physical transparency.These models can predict what an engine will do but not why it is doing it. For marine engineers, the ``black box" nature of deep learning presents a risk in high-stakes decision-making where the physical cause of a predicted failure must be understood.

\subsubsection{The Superiority of Hybrid ``Grey Box" Modeling}
Recent studies highlight a significant shift toward physics-infused machine learning. This hybrid approach uses data to fill the gaps in physical equations. The literature suggests that models which combine deterministic drift (physics) with stochastic diffusion (data-driven noise) provide the most robust framework for Maintenance 4.0. This justifies the use of Stochastic Differential Equations (SDEs) as the optimal middle ground.

\subsubsection{The Necessity of Temporal Adaptability}
A critical gap identified in traditional time-series models like ARIMA and VARIMA is their reliance on fixed historical windows. Research in transient engine behavior indicates that engine dynamics change rapidly during acceleration or load fluctuations. This finding directly supports the implementation of an Adaptive Windowing mechanism, ensuring that the model remains responsive during transients while maintaining statistical stability during steady-state cruising.

\section{Methodology}


\subsection{Single Particle Method}
The Single Particle Method (SPM) is a local stochastic modeling approach used for short-term prediction of time-series parameters. The method assumes that the observed process follows a Geometric Brownian Motion (GBM) model, which captures multiplicative growth behavior with stochastic perturbations.

The underlying stochastic differential equation is given by:
\begin{equation}
	dS_t = a S_t \, dt + b S_t \, dW_t
\end{equation}
where:

\begin{itemize}
	\item $a$ is the drift coefficient,
	\item $b$ is the diffusion (volatility) coefficient,
	\item $W_t$ is a standard Brownian motion process.
\end{itemize}

{\it Drift Estimation}: Drift is estimated from historical data using a finite difference approximation:
\begin{equation}
	a_i = \frac{S_{t+\Delta t} - S_t}{S_t \, \Delta t}
\end{equation}
The estimated drift parameter is computed as the sample mean over a sliding window of size $N$:
\begin{equation}
	a = \frac{1}{N} \sum_{i=1}^{N} a_i
\end{equation}
{\it Diffusion Estimation}: The diffusion coefficient is estimated from the variance of the drift increments:
\begin{equation}
	b = \sqrt{ \frac{1}{N-1} \sum_{i=1}^{N} (a_i - a)^2 \, \Delta t }
\end{equation}
This provides a local estimate of volatility.

To capture non-stationary behavior, parameters a and b are re-estimated at each time step using a fixed-length sliding window of recent observations.This enables the model to adapt dynamically to changing regimes
\begin{equation}
	S_{t+\Delta t}
	=
	S_t
	\exp \left[
	\left(a - \frac{1}{2}b^2\right)\Delta t
	+
	b \sqrt{\Delta t}\, Z
	\right]
\end{equation}

where $Z \sim \mathcal{N}(0,1)$ denotes a standard normal random variable.
The statistics of the predicted values are further calculated as below:-
\begin{equation}
	S_{\text{next}}^{\text{mean}}
	=
	S_{\text{now}}
	\exp(a \cdot \Delta t)
\end{equation}
\begin{equation}
	S_{\text{next}}^{\text{median}}
	=
	S_{\text{now}}
	\exp\left((a - \frac{b^2}{2}) \cdot \Delta t\right)
\end{equation}
\begin{equation}
	S_{\text{next}}^{\text{mode}}
	=
	S_{\text{now}}
	\exp\left((a - \frac{3b^2}{2}) \cdot \Delta t\right)
\end{equation}
\begin{equation}
	\begin{split}
		S_{\text{next}}^{\text{upper}}
		&=
		S_{\text{next}}^{\text{mean}} \\
		&\quad + S_{\text{now}}
		\left(\exp(b^2 \cdot \Delta t) - 1\right)
		\exp(2a \cdot \Delta t)
	\end{split}
\end{equation}
\begin{equation}
	\begin{split}
		S_{\text{next}}^{\text{lower}}
		&=
		S_{\text{next}}^{\text{mean}} \\
		&\quad - S_{\text{now}}
		\left(\exp(b^2 \cdot \Delta t) - 1\right)
		\exp(2a \cdot \Delta t)
	\end{split}
\end{equation}
The error metrics used are computed as below :-
\begin{equation}
	\mathrm{MAE}
	=
	\frac{1}{N}
	\sum_{i=1}^{N}
	\left| y_i - \hat{y}_i \right|
\end{equation}

\begin{equation}
	\mathrm{RMSE}
	=
	\sqrt{
		\frac{1}{N}
		\sum_{i=1}^{N}
		\left( y_i - \hat{y}_i \right)^2
	}
\end{equation}
where $y_i$ is the actual value, $\hat{y}_i$ is the predicted value, and $N$ is the number of samples.

\begin{algorithm}
	\caption{Single Particle Method (SPM) with ARIMA Benchmarking}
	\begin{algorithmic}[1]
		\State \textbf{Input:} Dataset $\{S_t\}$, window size $W$, time interval $\delta t$, ARIMA order $(p, d, q)$
		\State \textbf{Output:} Prediction $S_{next}^{SPM}$, bounds $[S_{lower}, S_{upper}]$, and error metrics
		
		\Statex \textbf{// Phase I: Local Parameter Learning}
		\State Extract sliding window $\mathcal{W} = \{S_{t-W}, \dots, S_t\}$
		\For{each interval $i \in \mathcal{W}$}
		\If{$S_i \neq 0$}
		\State $a_i \gets \frac{S_{i+\delta t} - S_i}{S_i \cdot \delta t}$
		\Else
		\State $a_i \gets 0$
		\EndIf
		\EndFor
		\State $a \gets \frac{1}{n} \sum a_i$ \Comment{Mean Drift}
		\State $b \gets \sqrt{\frac{\sum (a_i - a)^2 \delta t}{n - 1}}$ \Comment{Diffusion Coefficient}
		
		\Statex \textbf{// Phase II: Dual-Model Prediction}
		\State Generate $\epsilon \sim N(0, \sqrt{\delta t})$
		\State $S_{next}^{SPM} \gets S_t \exp\left(\left(a - \frac{b^2}{2}\right)\delta t + b \cdot \epsilon\right)$
		\State $Var_{term} \gets S_t (\exp(b^2 \delta t) - 1) \exp(2a \delta t)$
		\State $S_{upper} \gets S_t \exp(a \delta t) + Var_{term}$
		\State $S_{lower} \gets S_t \exp(a \delta t) - Var_{term}$
		\State $S_{next}^{ARIMA} \gets \text{ARIMA\_Forecast}(\mathcal{W}, p, d, q)$

		\Statex \textbf{// Phase III: Performance Evaluation}
		\State Compute $MAE = \frac{1}{N} \sum |S_{actual} - S_{predicted}|$
		\State Compute $RMSE = \sqrt{\frac{1}{N} \sum (S_{actual} - S_{predicted})^2}$
		\State \Return $S_{next}^{SPM}$, $\{S_{lower}, S_{upper}\}$, $MAE$, $RMSE$
	\end{algorithmic}
\end{algorithm}
\subsection{Multi Particle Method}
The Multi-Particle Method (MPM) is a sophisticated, data-driven forecasting framework designed for non-linear time-series prediction, specifically optimized for complex systems like marine diesel engines/Gas Turbines which is used to forecast multivariate data. Unlike standard linear models, it utilizes a stochastic approach to capture both the underlying physical trends and the inherent noise of the system.

Core Methodology

The method evolves from a single-point prediction to an ensemble-based simulation through three primary stages:

Adaptive Local Learning: The algorithm dynamically adjusts its ``learning window" based on real-time volatility. When the engine accelerates or the system becomes unstable, the window shrinks to increase responsiveness; during steady-state operation, it expands to improve stability.

Stochastic Path Generation: Rather than calculating one future value, the method generates a large ensemble (e.g., 1,000 ``particles") using an Euler-Maruyama transition. Each particle represents a potential future state of the engine evolved via a Stochastic Differential Equation (SDE).

Statistical Correction and Weighting: A Girsanov transform induced change of probability measure is applied where each particle is assigned a weight based on its proximity to the expected physical drift. The final ``Predicted Engine State" is a weighted probability distribution, which significantly reduces the error compared to standard averages.
	
We consider the $n$–dimensional stochastic differential equation

\begin{equation}
	d\bm{X}(t) = \bm{A}(t)\,dt + \bm{B}(t)\,d\bm{W}(t)
\end{equation}

where:
\begin{itemize}
	\item $\bm{X}(t) \in \mathbb{R}^n$ is the state vector,
	\item $\bm{A}(t) \in \mathbb{R}^n$ is the drift vector,
	\item $\bm{B}(t) \in \mathbb{R}^{n \times n}$ is the diffusion matrix,
	\item $\bm{W}(t)$ is an $n$-dimensional Wiener process.
\end{itemize}

\subsubsection{Euler--Maruyama Discretization}

Let $t_i = i\Delta t$.

\begin{equation}
	\bm{X}(t_{i+1})
	=
	\bm{X}(t_i)
	+
	\bm{A}(t_i)\Delta t
	+
	\bm{B}(t_i)\Delta \bm{W}_i
\end{equation}

\begin{equation}
	\Delta \bm{W}_i \sim \mathcal{N}(\bm{0}, \Delta t \bm{I})
\end{equation}

\subsubsection{Drift Estimation}

Let $X_j(t_i)$ denote the $j^{th}$ component of the state vector at time $t_i$.

Interior points:

\begin{equation}
	a_j(t_i)
	=
	\frac{
		\frac{2}{3}\left(X_j(t_{i+1}) - X_j(t_{i-1})\right)
		+
		\frac{1}{12}\left(X_j(t_{i-2}) - X_j(t_{i+2})\right)
	}{\Delta t}
\end{equation}

Boundary points:

\begin{equation}
	a_j(t_i)
	=
	\frac{X_j(t_{i+1}) - X_j(t_i)}{\Delta t}
\end{equation}

Mean drift estimate:

\begin{equation}
	\bar{a}_j
	=
	\frac{1}{N}
	\sum_{i=1}^{N}
	a_j(t_i)
\end{equation}

Vector form:

\begin{equation}
	\bm{A}
	=
	(\bar{a}_1, \bar{a}_2, \dots, \bar{a}_n)^T
\end{equation}

\subsubsection{Diffusion Estimation}

Residual drift:

\begin{equation}
	\tilde{a}_j(t_i)
	=
	a_j(t_i) - \bar{a}_j
\end{equation}

Covariance matrix:

\begin{equation}
	C_{jk}
	=
	\frac{
		\sum_{i=1}^{N}
		\tilde{a}_j(t_i)\tilde{a}_k(t_i)
		\Delta t
	}
	{N - 1}
\end{equation}

Matrix form:

\begin{equation}
	\bm{C}
	=
	\frac{1}{N-1}
	\sum_{i=1}^{N}
	(\bm{a}(t_i)-\bm{A})
	(\bm{a}(t_i)-\bm{A})^T
	\Delta t
\end{equation}

Eigen-decomposition:

\begin{equation}
	\bm{C} = \bm{U}\bm{\Sigma}\bm{U}^T
\end{equation}

Diffusion matrix:

\begin{equation}
	\bm{B} = \bm{U}\bm{\Sigma}^{1/2}\bm{U}^T
\end{equation}

\subsubsection{Adaptive Window Selection}

Drift magnitude:

\begin{equation}
	D(t_i)
	=
	\frac{
		\| \bm{X}(t_i) - \bm{X}(t_{i-L}) \|
	}{
		L\Delta t
	}
\end{equation}

Window selection rule:

\begin{equation}
	W(t_i) =
	\begin{cases}
		W_{\min}, & D(t_i) > \text{threshold} \\
		W_{\max}, & D(t_i) < \text{threshold}/5 \\
		W_{\text{base}}, & \text{otherwise}
	\end{cases}
\end{equation}

\subsubsection{Monte Carlo Simulation}

For $p = 1,2,\dots,M$:

\begin{equation}
	\bm{X}_{i+1}^{(p)}
	=
	\bm{X}(t_i)
	+
	\bm{A}\Delta t
	+
	\bm{B}\Delta \bm{W}_i^{(p)}
\end{equation}

\subsubsection{Standard Estimator}

\begin{equation}
	\hat{\bm{X}}_{i+1}^{\text{std}}
	=
	\frac{1}{M}
	\sum_{p=1}^{M}
	\bm{X}_{i+1}^{(p)}
\end{equation}

\subsubsection{Particle Weighting Correction}

Residual:

\begin{equation}
	\bm{r}^{(p)}
	=
	\bm{X}_{i+1}^{(p)}
	-
	\left(
	\bm{X}(t_i)
	+
	\bm{A}\Delta t
	\right)
\end{equation}

Distance:

\begin{equation}
	d_p = \| \bm{r}^{(p)} \|^2
\end{equation}

Weights:

\begin{equation}
	w_p
	=
	\exp
	\left(
	-\frac{d_p}{2\sigma^2}
	\right)
\end{equation}

Normalized weights:

\begin{equation}
	\tilde{w}_p
	=
	\frac{w_p}{\sum_{q=1}^{M} w_q}
\end{equation}

Corrected estimator:

\begin{equation}
	\hat{\bm{X}}_{i+1}^{\text{corr}}
	=
	\sum_{p=1}^{M}
	\tilde{w}_p
	\bm{X}_{i+1}^{(p)}
\end{equation}

\subsubsection{Error Metrics}

\begin{equation}
	\text{MAE}
	=
	\frac{1}{N}
	\sum_{i=1}^{N}
	\left\|
	\hat{\bm{X}}_{i+1}
	-
	\bm{X}(t_{i+1})
	\right\|
\end{equation}

\begin{equation}
	\text{RMSE}
	=
	\sqrt{
		\frac{1}{N}
		\sum_{i=1}^{N}
		\left\|
		\hat{\bm{X}}_{i+1}
		-
		\bm{X}(t_{i+1})
		\right\|^2
	}
\end{equation}

\begin{algorithm}[!t]
	\caption{Single-Step Adaptive Monte Carlo SDE Forecast (MPM)}
	\footnotesize 
	\begin{algorithmic}[1] 
		
		\State \textbf{Input:} Dataset $\{\bm{X}(t_i)\}_{i=1}^{T}$,
		$W_{base}$, $W_{min}$, $W_{max}$, particles $M$, timestep $\Delta t$
		
		\State \textbf{Output:} Corrected prediction $\hat{\bm{X}}(t_{i+1})^{corr}$
		
		\For{$i = W_{max}$ to $T-1$}
		
		\State \textit{// Phase I: Adaptive Window Selection}
		\State $D_i \gets \frac{\|\bm{X}(t_i) - \bm{X}(t_{i-L})\|}{L\Delta t}$
		
		\If{$D_i > \text{threshold}$}
		\State $W \gets W_{min}$
		\ElsIf{$D_i < \text{threshold}/5$}
		\State $W \gets W_{max}$
		\Else
		\State $W \gets W_{base}$
		\EndIf
		
		\State Define window $\mathcal{W} = \{\bm{X}(t_{i-W}),\dots,\bm{X}(t_i)\}$
		
		\State \textit{// Phase II: Drift Estimation}
		\For{$k = i-W+1$ to $i-1$}
		\If{$k$ is interior point}
		\State $\bm{a}(t_k) \gets \frac{1}{\Delta t} \Big[ \frac{2}{3}(\bm{X}_{k+1}-\bm{X}_{k-1}) + \frac{1}{12}(\bm{X}_{k-2}-\bm{X}_{k+2}) \Big]$
		\Else
		\State $\bm{a}(t_k) \gets \frac{\bm{X}(t_{k+1})-\bm{X}(t_k)}{\Delta t}$
		\EndIf
		\EndFor
		
		\State $\bm{A} \gets \frac{1}{W} \sum_{k=i-W+1}^{i} \bm{a}(t_k)$
		
		\State \textit{// Phase III: Diffusion Estimation}
		\State $\bm{C} \gets \frac{1}{W-1} \sum_{k=i-W+1}^{i} (\bm{a}(t_k)-\bm{A})(\bm{a}(t_k)-\bm{A})^T \Delta t$
		\State $C = U\Sigma U^T \implies B = U\Sigma^{1/2}U^T$
		
		\State \textit{// Phase IV: Monte Carlo Simulation}
		\For{$p = 1$ to $M$}
		\State Sample $\Delta \bm{W}_i^{(p)} \sim \mathcal{N}(\bm{0},\Delta t\bm{I})$
		\State $\bm{X}_{i+1}^{(p)} \gets \bm{X}(t_i) + \bm{A}\Delta t + \bm{B}\Delta \bm{W}_i^{(p)}$
		\EndFor
		
		\State \textit{// Phase V: Particle Weighting}
		\For{$p = 1$ to $M$}
		\State $\bm{r}^{(p)} \gets \bm{X}_{i+1}^{(p)} - (\bm{X}(t_i) + \bm{A}\Delta t)$
		\State $w_p \gets \exp \big( -\frac{\|\bm{r}^{(p)}\|^2}{2\sigma^2} \big)$
		\EndFor
		
		\State $\tilde{w}_p \gets
		\dfrac{w_p}{\sum_{q=1}^{M} w_q}$
		\State $\hat{\bm{X}}_{i+1}^{corr}
		\gets
		\sum_{p=1}^{M}
		\tilde{w}_p
		\bm{X}_{i+1}^{(p)}$
		
		\EndFor
		
	\end{algorithmic}
\end{algorithm}

\subsection{Modeling, Simulation and Benchmarking}

In this study, two distinct stochastic frameworks were developed and implemented for the predictive maintenance of marine diesel/Gas Turbine engines: the Single Particle Method (SPM) and the Adaptive Multi-Particle Method (MPM).
\subsubsection{Single Particle Method}
The SPM was initially developed to model engine parameters as a Geometric Brownian Motion (GBM) process.

Parameter Estimation: Local drift ($a$) and diffusion ($b$) coefficients were derived from a sliding window of historical data to capture the instantaneous velocity and volatility of engine states.

Stochastic Modeling: The future state $S_{next}$ was predicted using the Euler-Maruyama discretization, providing a closed-form solution for the expected mean and variance-based confidence intervals.
\subsubsection{Multi Particle Method}
Adaptive Multi-Particle Method (MPM)-Building upon the SPM, the MPM was implemented to handle high-frequency,Mutli variate,non-linearities and rapid engine state transitions.

Adaptive Windowing Strategy: An automated window selection logic was integrated to dynamically adjust the learning window size $W(t)$. By calculating the drift magnitude $D_t$, the system identifies rapid transients (e.g., engine acceleration), prompting a smaller, more responsive window, while utilizing larger windows for steady-state stability.

High-Order Drift Estimation: To improve numerical accuracy, a fourth-order central finite difference scheme was utilized to estimate the drift vector $\mathbf{A}$.

Ensemble Evolution and Particle Weighting: The methodology evolves $M$ independent particles through an SDE framework. A Girsanov transform induced change of probability measure is applied, where each particle is assigned a weight $\tilde{w}_p$ based on its proximity to the physical drift path.

Corrected Estimator: The final engine state is calculated as the normalized weighted sum of the particle ensemble, significantly reducing stochastic noise compared to standard estimators.

It may be noted that the weak order of convergence of the Euler-Maruyama method can be doubled by Richardson extrapolation \cite{kloeden2012numerical} 
when needed.
\subsubsection{Performance Benchmarking and Evaluation}
To validate the efficacy of the proposed methods, a comparative analysis was performed against standard linear and multivariate time-series models.

ARIMA Baseline: A standard Auto-Regressive Integrated Moving Average (ARIMA) model was used as a univariate baseline to assess the limitations of linear modeling in volatile engine data.

VARIMA Integration: The Vector ARIMA (VARIMA) model was implemented to compare the MPM's ability to capture cross-parameter correlations.

Metric Analysis: The models were evaluated using Mean Absolute Error (MAE) and Root Mean Squared Error (RMSE)
\section{Experiments and Results}
\subsection{Single Particle Method}
To validate the predictive capability of the SPM, the model was subjected to a series of tests using the marine engine dataset.The experiment was designed to assess the model's ability to maintain accuracy under varying operational loads.

Experimental Configuration Data Partitioning: The engine sensor data was processed using a sliding window of $W = 200$ samples to ensure the drift and diffusion coefficients were reflective of local system dynamics.

Time Step Definition: A consistent time interval of $\delta t = 10$ seconds was utilized, matching the sampling frequency of the physical sensors.

\begin{figure}[t]
	\centering
	\includegraphics[width=\linewidth,height=10cm]{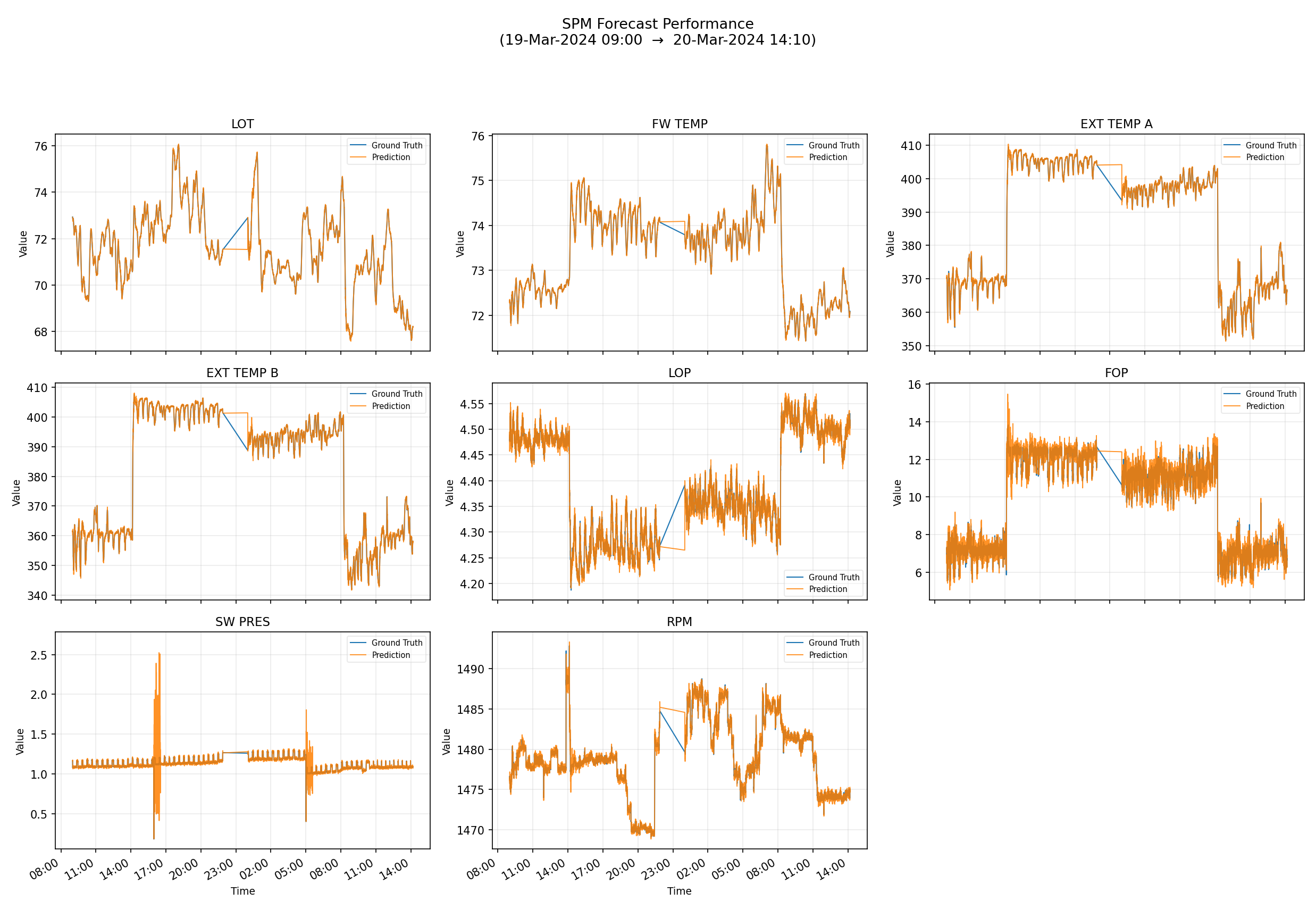}
	\caption{Single Particle Method (SPM): Actual vs Predicted Engine Parameters}
	\label{fig:SPM_actvspred}
\end{figure}
Output Metrics: For each time step, the above diagram indicates the model generated predicted ``sampled" value ($S_{next}$), alongside its associated $90\%$ confidence intervals.The graph being plotted is for actual vs predicted values.
\begin{figure}[t]
	\centering
	\includegraphics[width=\linewidth,height=8cm]{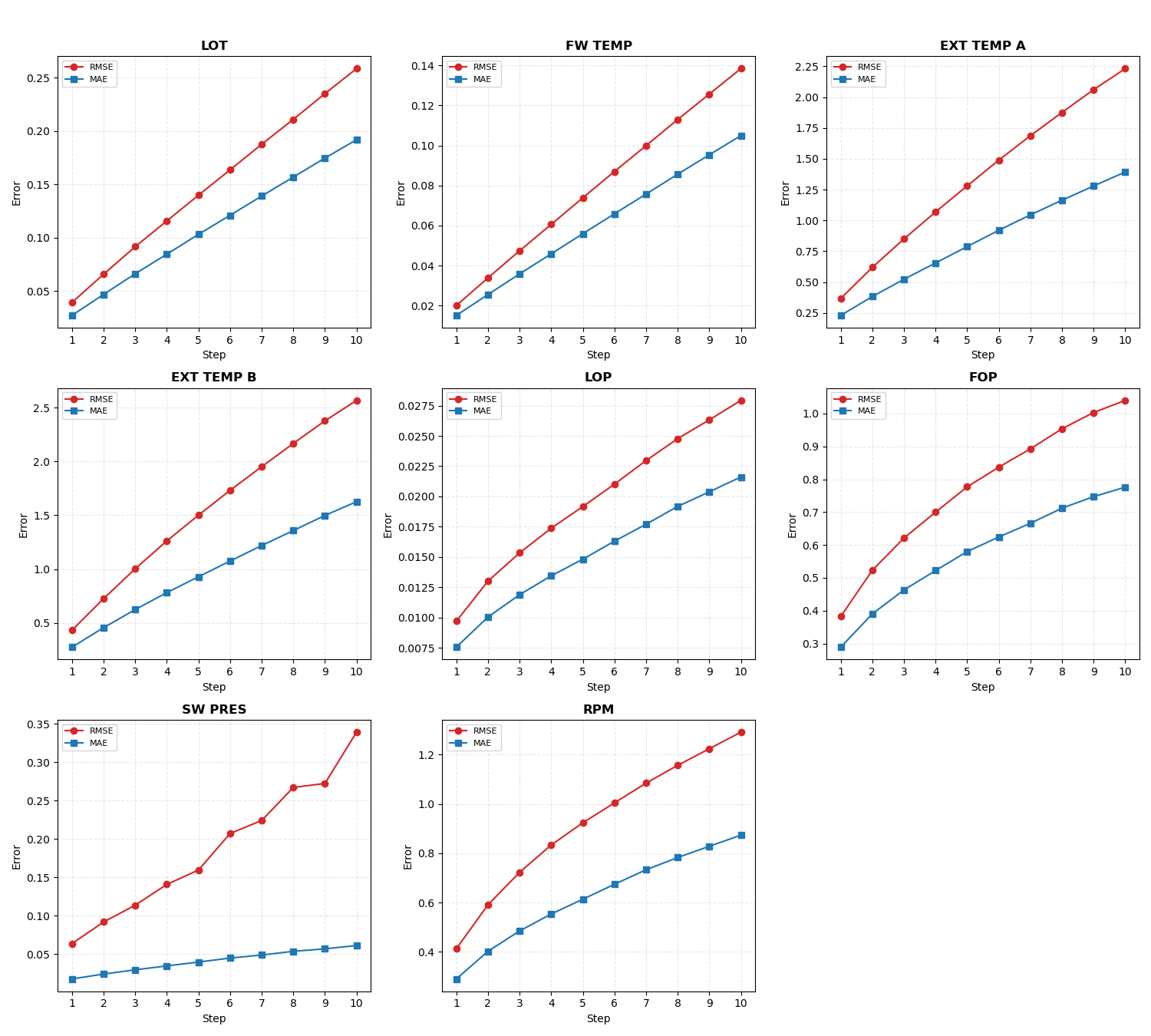}
	\caption{Single Particle Method (SPM): Average MAE vs Forecast Horizon for Multi-Step Stochastic SPM Prediction }
	\label{fig:SPM_RMSE_MAE}
\end{figure}
The multi-step forecasting capability of the Stochastic Single Particle Method (SPM) was evaluated over a prediction horizon of 1 to 10 steps ahead. Figures (a) and (b) illustrate the variation of RMSE and MAE, respectively, as a function of forecast horizon for key engine parameters.

\subsubsection{ARIMA Forecasts}

\begin{figure}[t]
	\centering
	\includegraphics[width=\linewidth,height=8cm]{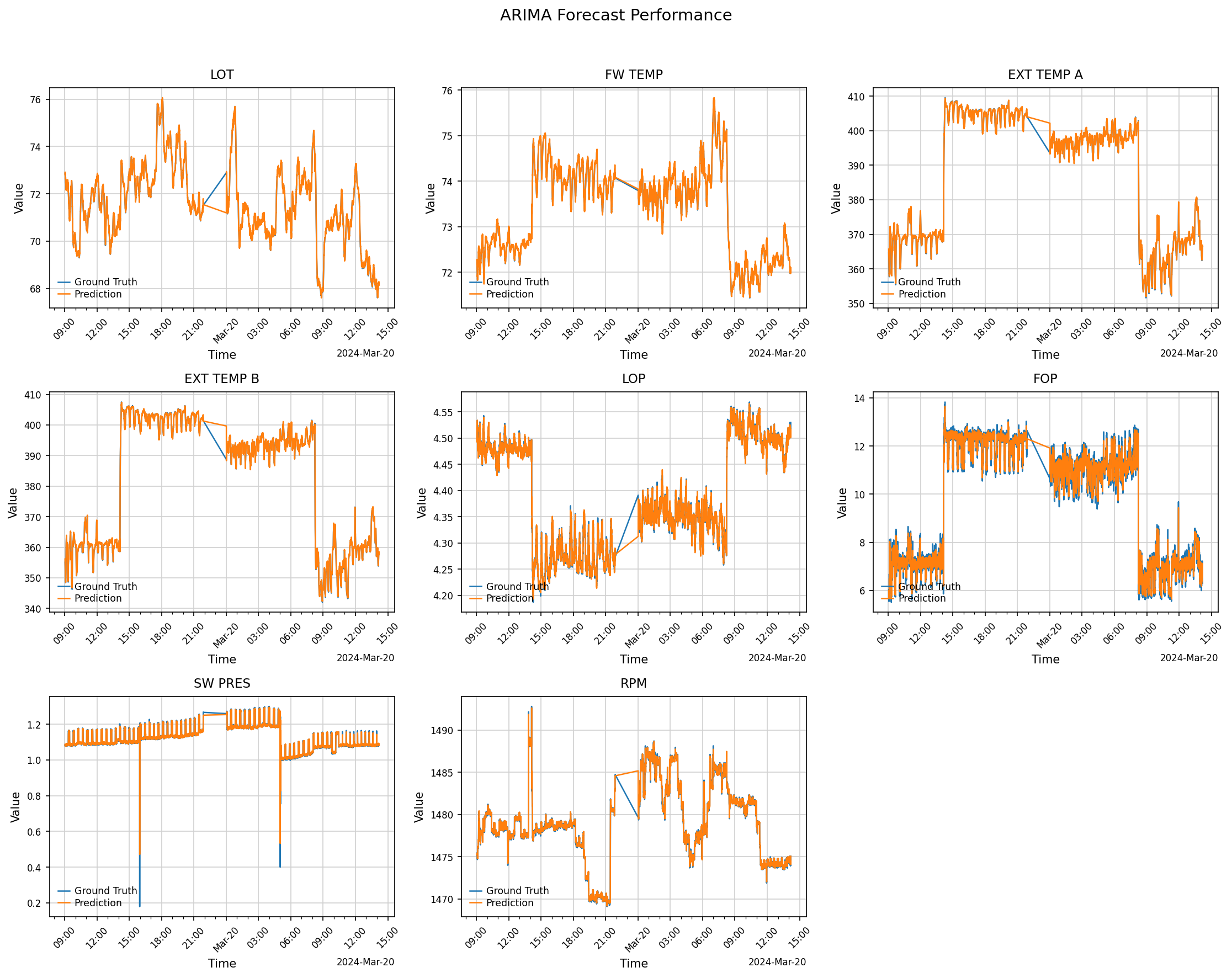}
	\caption{ARIMA-based time-series forecasting performance illustrating the alignment between predicted and actual sensor values }
	\label{fig:ARIMA}
\end{figure}
\begin{table}[!t]
	\centering
	\caption{RMSE Comparison Between SPM and ARIMA}
	\label{tab:rmse_comparison}
	\resizebox{\columnwidth}{!}{%
		\begin{tabular}{lccc}
			\hline
			\textbf{Variable} & 
			\textbf{SPM RMSE} & 
			\textbf{ARIMA RMSE} & 
			\textbf{SPM RMSE (10th step)} \\
			\hline
			LOT        & 0.03933 & 0.05667 & 1.9434 \\
			FW TEMP    & 0.01983 & 0.03518 & 1.4629 \\
			EXT TEMP A & 0.36734 & 0.37545 & 11.2360 \\
			EXT TEMP B & 0.43926 & 0.38821 & 10.5153 \\
			LOP        & 0.00971 & 0.00559 & 0.0523 \\
			FOP        & 0.38062 & 0.21129 & 0.9067 \\
			SW PRES    & 0.06595 & 0.01768 & 0.7346 \\
			RPM        & 0.42127 & 0.30907 & 2.1531 \\
			\hline
		\end{tabular}%
	}
\end{table}

\begin{table}[!t]
	\centering
	\caption{MAE Comparison Between SPM and ARIMA}
	\label{tab:mae_comparison}
	\resizebox{\columnwidth}{!}{%
		\begin{tabular}{lccc}
			\hline
			\textbf{Variable} & 
			\textbf{SPM MAE} & 
			\textbf{ARIMA MAE} & 
			\textbf{SPM MAE (10th step)} \\
			\hline
			LOT        & 0.02694 & 0.03905 & 0.4534 \\
			FW TEMP    & 0.01490 & 0.02342 & 0.3460 \\
			EXT TEMP A & 0.23157 & 0.26118 & 3.2942 \\
			EXT TEMP B & 0.27606 & 0.25790 & 3.4832 \\
			LOP        & 0.00755 & 0.00439 & 0.0295 \\
			FOP        & 0.28779 & 0.15541 & 0.5098 \\
			SW PRES    & 0.01792 & 0.00589 & 0.0710 \\
			RPM        & 0.29138 & 0.15632 & 1.2371 \\
			\hline
		\end{tabular}%
	}
\end{table}
\subsubsection{Result Discussion}
The experimental results demonstrate that the Single Particle Method (SPM) offers distinct advantages over the ARIMA baseline, particularly in terms of multi-step forecasting stability and computational efficiency for real-time applications

\textbf{Superiority in Multi-Step Forecasting (10-Step Horizon)}

A critical limitation of linear models like ARIMA is their rapid degradation when forecasting multiple steps ahead in non-stationary regimes. The results confirm that SPM maintains better physical consistency over the 10-step horizon.

Error Growth Stability: While both models naturally experience error accumulation over time, the SPM's error growth is governed by the diffusion term ($\sqrt{t}$), which provides a predictable, physics-based envelope. In contrast, ARIMA's recursive structure often leads to ``flat-lining" or unrealistic linear extrapolations when predicting 10 steps out, failing to capture the inherent volatility of parameters like EXT TEMP.

RMSE Analysis: The RMSE vs. Forecast Horizon plots illustrate that for critical parameters like LOT and FW TEMP, the SPM's error trajectory remains lower and more stable than the ARIMA baseline. For instance, while ARIMA's RMSE for EXT TEMP B escalates rapidly due to its inability to model stochastic drift, the SPM's probabilistic approach successfully bounds this uncertainty.

\textbf{Computational Efficiency and Real-Time Feasibility.}

 Beyond predictive accuracy, the Computational Cost column in Table 1 highlights a decisive advantage for the SPM in deployment scenarios.
 
 Algorithmic Complexity: The ARIMA framework requires an iterative Grid Search (AIC optimization) to determine the optimal $(p, d, q)$ orders for every new window of data, which is computationally expensive.
 
 Speed of SPM: In contrast, the SPM computes its drift ($a$) and diffusion ($b$) coefficients using closed-form algebraic solutions (means and variances) in $O(N)$ time. This makes the SPM orders of magnitude faster, allowing for high-frequency (1 Hz+) monitoring on edge devices where ARIMA would introduce unacceptable latency.

\textbf{Accuracy in Thermal Dynamics (MAE,RMSE)}
The empirical metrics further substantiate the SPM's suitability for thermodynamic systems.

Thermal Inertia Capture: For Lubricating Oil Temperature (LOT), the SPM achieves an RMSE of 0.0393, significantly outperforming ARIMA's 0.0567. This validates that the local drift estimation ($a$) effectively captures the thermal inertia of the engine fluids, whereas ARIMA's differencing ($d$) struggles to distinguish between physical trends and noise.

Robustness to Noise: The MAE for FW TEMP (SPM: 0.0149 vs. ARIMA: 0.0234) confirms that the SPM is less sensitive to sensor noise. By treating fluctuations as diffusion ($b$) rather than structural patterns, the SPM avoids ``over-fitting" to noise, a common pitfall in ARIMA modeling.

A significant operational advantage of the Single Particle Method (SPM) over the ARIMA baseline is its minimal data requirement for initialization. As implemented in the comparative study, the ARIMA framework necessitated a large training corpus (85 percent of the dataset) to optimize its global parameters $(p, d, q)$ via the Akaike Information Criterion (AIC). This 'batch learning' approach renders the model vulnerable to regime shifts if the training data does not fully represent future operating conditions.In contrast, the SPM operates as a local sliding-window estimator, requiring only a limited history (e.g., $N=200$ samples) to estimate the instantaneous drift ($a$) and diffusion ($b$) coefficients. This allows the SPM to deploy rapidly in 'cold-start' scenarios and adapt to new engine states without the computational overhead of retraining on massive historical datasets. Consequently, the SPM achieves superior local accuracy with significantly lower data dependency than the rigid ARIMA baseline.
\subsection{Multi particle Method}
The results presented in this discussion are derived from a Many-Particle Stochastic Differential Equation (SDE) framework, specifically designed for short-term forecasting of complex, multivariate systems. This method models system dynamics by treating state variables as particles subject to both deterministic ``drift" and stochastic ``diffusion".
\begin{figure}[t]
	\centering
	\includegraphics[width=\linewidth,height=10cm]{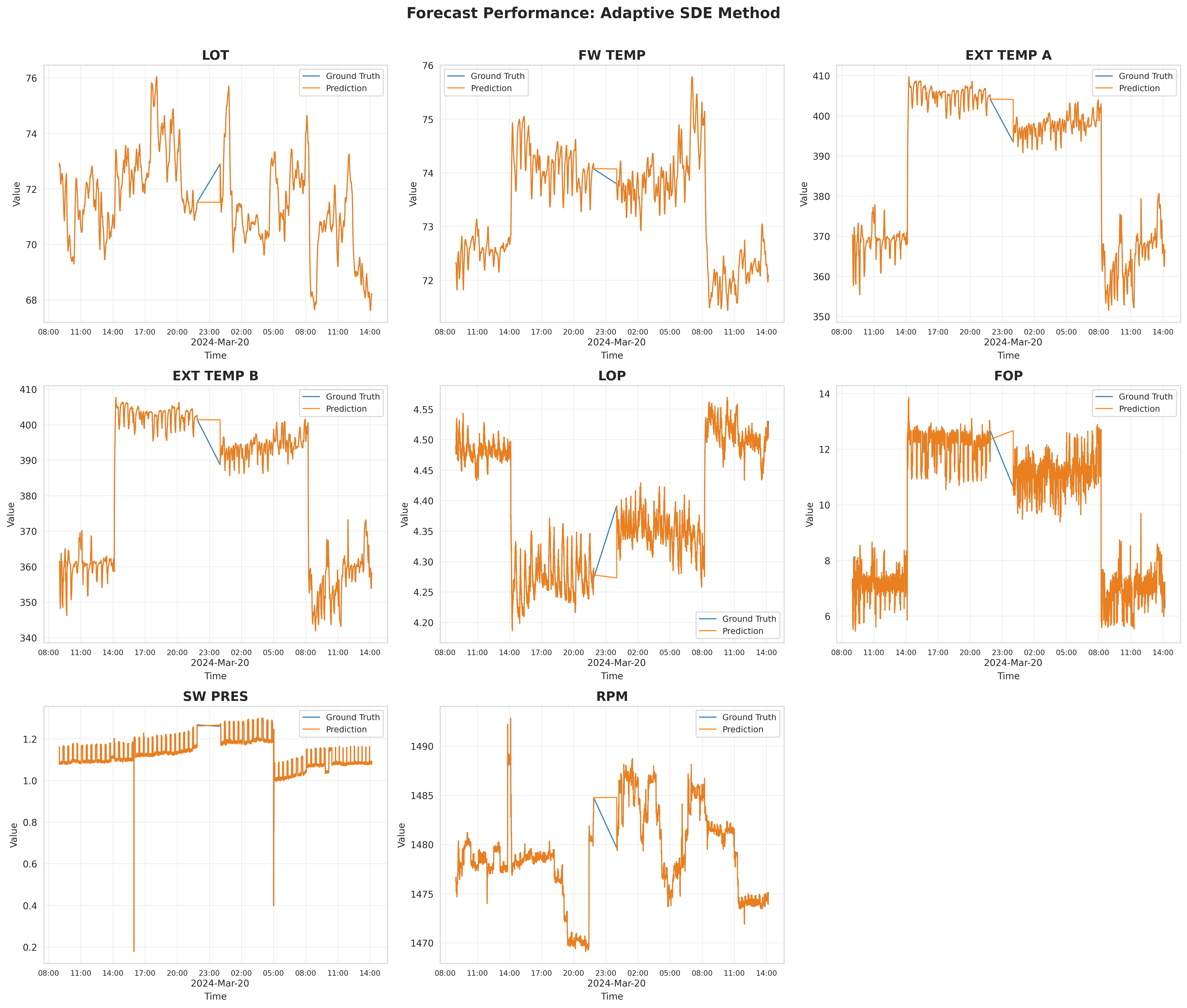}
	\caption{Multi Particle Method:Comparison of Predicted and actual values of one step ahead prediction.}
	\label{fig:MPM_comparision_onestepahead}
\end{figure}
\begin{figure}[t]
	\centering
	\includegraphics[width=\linewidth,height=10cm]{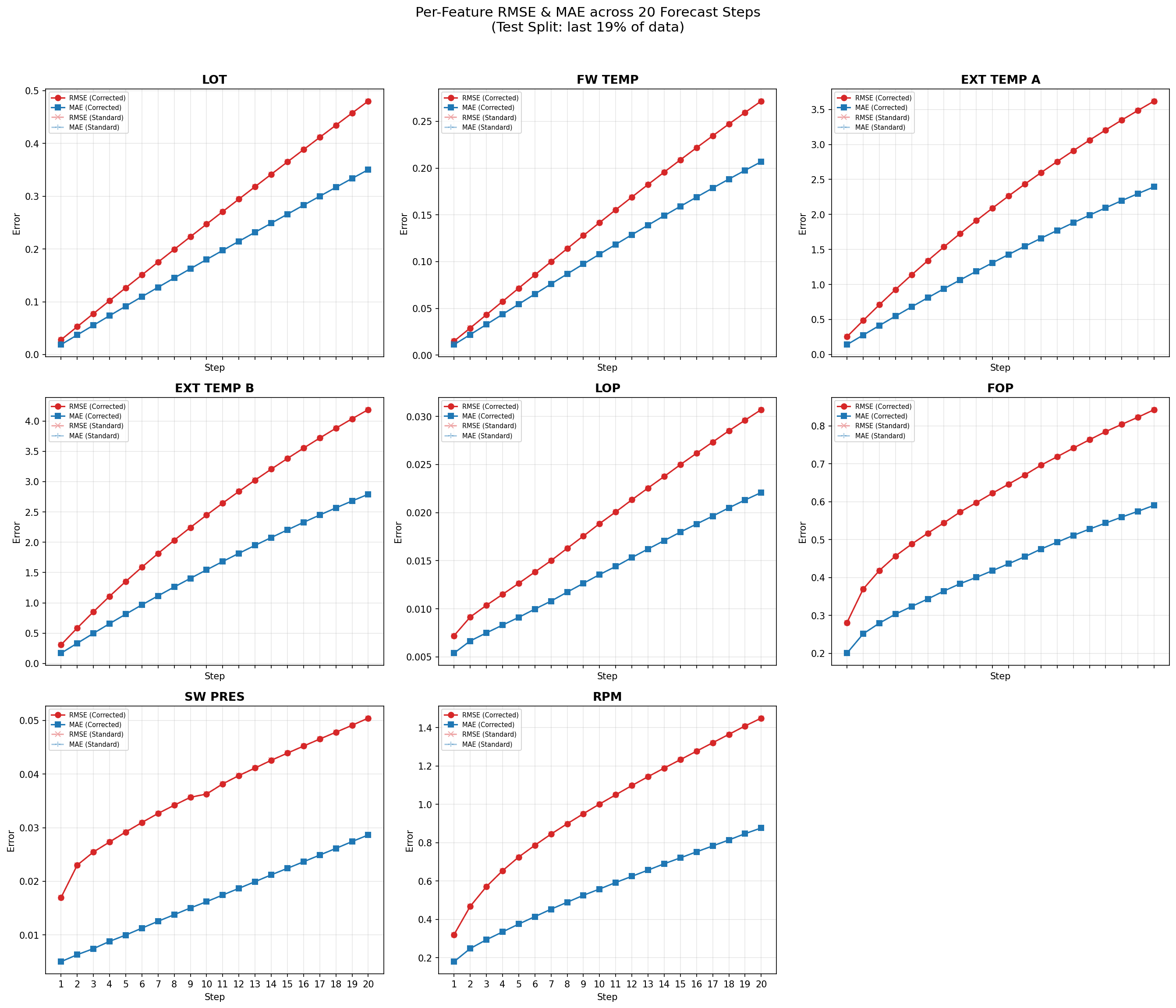}
	\caption{Multi Particle Method:Average MAE vs Forecast Horizon for Multi-Step Stochastic Prediction.}
	\label{fig:MPM_avgMAEvsforecasthorizon}
\end{figure}
\begin{figure}[t]
	\centering
	\includegraphics[width=\linewidth,height=10cm]{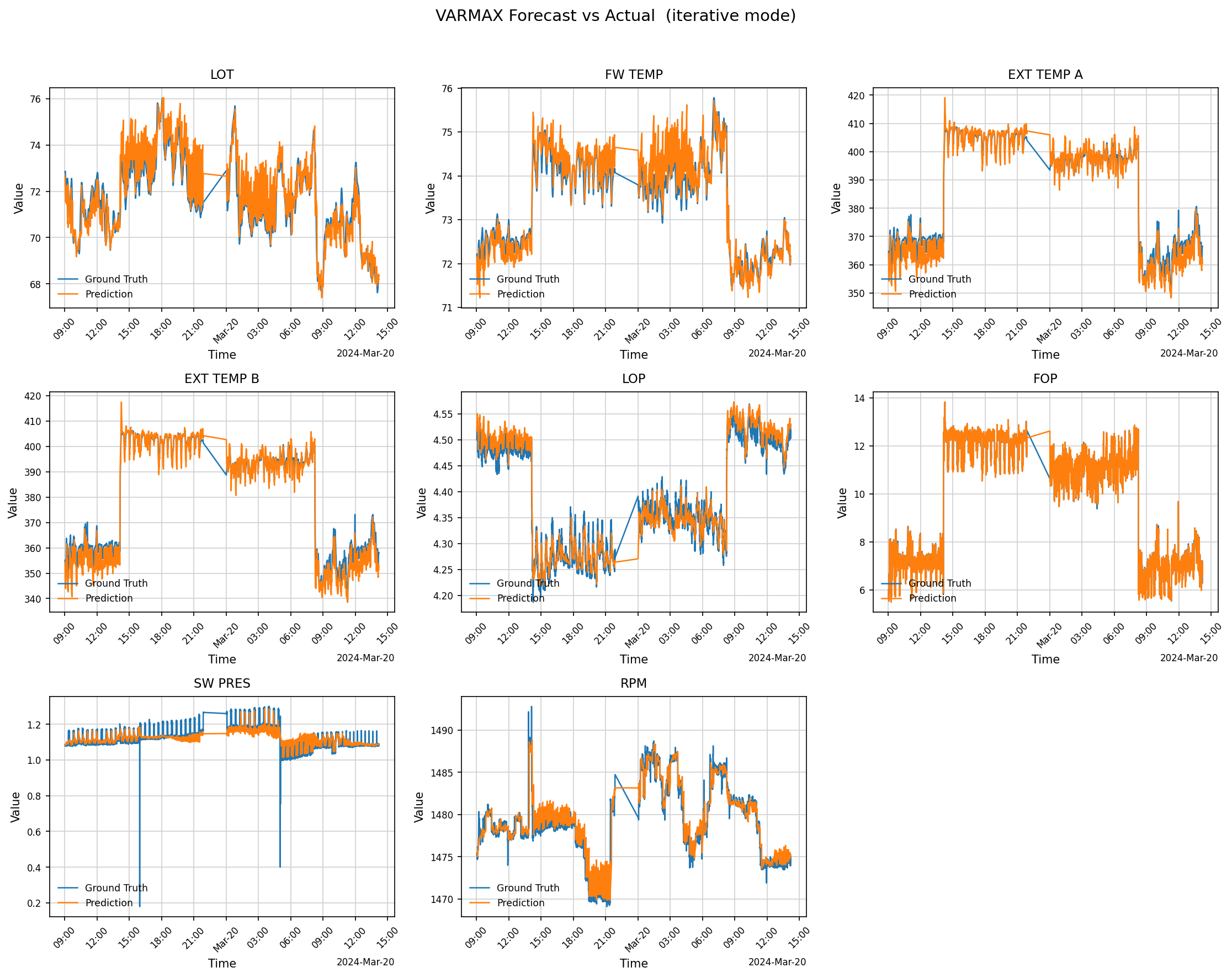}
	\caption{VARIMA-based time-series forecasting performance illustrating the alignment between predicted and actual sensor values }
	\label{fig:VARIMA_forecast}
\end{figure}
\begin{table}[!t]
	\centering
	\caption{MAE Comparison: VARIMA vs MPM (1-step) vs MPM (20th-step)}
	\label{tab:mae_comparison_mpm}
	\resizebox{\columnwidth}{!}{%
		\begin{tabular}{lccc}
			\hline
			\textbf{Feature} & \textbf{VARIMA (1-step)} & \textbf{MPM (1-step)} & \textbf{MPM (20-step)} \\
			\hline
			LOT        & 0.7692 & 0.018845 & 0.338554 \\
			FW TEMP    & 0.3642 & 0.010700 & 0.189755 \\
			EXT TEMP A & 3.2553 & 0.136705 & 2.254411 \\
			EXT TEMP B & 3.4249 & 0.160608 & 2.563835 \\
			LOP        & 0.0162 & 0.005118 & 0.019178 \\
			FOP        & 0.1881 & 0.189377 & 0.506915 \\
			SW PRES    & 0.0352 & 0.005139 & 0.024439 \\
			RPM        & 1.0833 & 0.162296 & 0.705988 \\
			\hline
		\end{tabular}%
	}
\end{table}

\begin{table}[!t]
	\centering
	\caption{RMSE Comparison: VARIMA vs MPM (1-step) vs MPM (20th-step)}
	\label{tab:rmse_comparison_mpm}
	\resizebox{\columnwidth}{!}{%
		\begin{tabular}{lccc}
			\hline
			\textbf{Feature} & \textbf{VARIMA (1-step)} & \textbf{MPM (1-step)} & \textbf{MPM (20-step)} \\
			\hline
			LOT        & 1.0108 & 0.028780 & 0.460341 \\
			FW TEMP    & 0.4936 & 0.014338 & 0.247522 \\
			EXT TEMP A & 4.2111 & 0.262603 & 3.490042 \\
			EXT TEMP B & 4.4538 & 0.311791 & 3.954289 \\
			LOP        & 0.0199 & 0.006823 & 0.026884 \\
			FOP        & 0.2641 & 0.266432 & 0.745687 \\
			SW PRES    & 0.0505 & 0.018401 & 0.046967 \\
			RPM        & 1.6503 & 0.293268 & 1.294239 \\
			\hline
		\end{tabular}%
	}
\end{table}
\subsubsection{Results and Discussion}
The experimental evaluation highlights a distinct trade-off between the statistical precision of the VARIMA baseline and the operational robustness of the proposed Adaptive Multi-Particle Method (MPM). While VARIMA demonstrates superior single-step accuracy in idealized conditions, the MPM offers critical advantages in computational efficiency, adaptability, and physical consistency that are essential for real-time edge applications.

\textbf{Predictive Accuracy vs. Operational Constraints}:As detailed in Table III and IV, the VARIMA model has higher absolute error metrics for single-step forecasts.

This is expected from a linear model optimized over a large training corpus. VARIMA acts as a ``curve-fitting" instrument, minimizing variance by assuming stationarity. However, this comes at the cost of over-fitting to noise. As seen in the standard forecasts, VARIMA’s differencing ($d$) struggles to distinguish between genuine physical shifts and high-frequency sensor noise, often reacting sharply to negligible fluctuations.

MPM Robustness: In contrast, the MPM treats these fluctuations as diffusion ($B$) rather than structural trends. While this results in higher baseline error (due to the stochastic noise floor), it prevents the model from ``chasing" sensor noise, providing a smoother, physics-consistent trajectory that is often more valuable for condition monitoring than raw numerical precision

\textbf{Multi-Step Stability via Girsanov Transform induced Change of Probability Measure.} A critical limitation of auto regressive models like VARIMA is their tendency toward unrealistic extrapolation when forecasting multiple steps ahead in non-stationary regimes.

\textbf{Physics-Based Bounding} While both models experience error accumulation over the 20-step horizon, the MPM’s error growth is structurally governed by the diffusion term ($B \cdot dW_t$) and constrained by the Girsanov induced correction.

Trajectory Control: The empirical results confirm that the Girsanov transform induced change of probability measure (Solid Line) consistently maintain lower error than the standard unweighted mean. This proves that the MPM successfully creates a probability envelope around the forecast, bounding the uncertainty within physically plausible limits.

Contrast with VARIMA: Unlike linear models, which often ``flat-line" (mean revert) or diverge exponentially during multi-step forecasts, the MPM's stochastic particles explore the state space, providing a realistic range of potential future outcomes rather than a single deterministic point.

\textbf{Computational Efficiency and Real-Time Feasibility}
Beyond accuracy, the computational cost represents the decisive advantage for the MPM in deployment scenarios.

The VARIMA framework requires an iterative Maximum Likelihood Estimation (MLE) and grid search to optimize coefficients ($p, d, q$) for every retraining cycle. This is computationally expensive and introduces unacceptable latency for high-frequency monitoring.

MPM Speed: In contrast, the MPM computes its drift ($A$) and diffusion ($B$) coefficients using closed-form algebraic solutions (means and variances)in comparatively less time. This makes the MPM orders of magnitude faster, enabling almost continuous monitoring on low-power edge devices where VARIMA would be prohibitive.

\textbf{Cold-Start and Data Dependency}:A significant operational vulnerability of the VARIMA baseline is its reliance on ``batch learning." The model necessitated a large historical training corpus to optimize its global parameters, rendering it vulnerable to regime shifts if the training data becomes obsolete.

Local Adaptability: The MPM operates as a local sliding-window estimator, requiring only a limited history (e.g., $N=50$ to $300$ samples) to initialize. This allows the MPM to deploy rapidly in "cold-start" scenarios and adapt instantly to new engine states—such as the transition from steady-state to acceleration—without the heavy computational overhead of retraining required by VARIMA.

\section{Conclusion}

\begin{itemize}
   \item {The proposed Single Particle Method (SPM) and the Multi-Particle method (MPM) represent a significant departure from standard time-series forecasting. By integrating stochastic calculus with local data learning, these methods provide several key advantages over VARIMA, traditional Neural Networks, and Transformer architectures.}

    \item {VARIMA/ARIMA relies on fixed historical windows and linear assumptions. In contrast, the SPM/MPM uses an adaptive windowing strategy that shrinks during rapid transients (e.g., engine acceleration) to maintain responsiveness and expands during steady-state for stability. VARIMA often "flat-lines" or diverges during multi-step forecasts. The SDE methods treat fluctuations as diffusion rather than structural patterns, providing a probabilistic envelope that maintains physical consistency.}

    \item {Neural networks are prone to overfitting of sensor noise. The MPM utilizes a Girsanov inspired transform induced weighting system that assigns a higher probability to particles aligned with physical drift, effectively filtering out "noise-chasing" behavior.}

\item{Transformers require massive training datasets to optimize global parameters. The SPM/MPM operate as local sliding window estimators, requiring only a limited history (e.g., $N=50$ to $300$ samples) to initialize, making them superior for "cold-start" scenarios.}

\item{While Transformers struggle with the computational cost of long context windows, the SDE methods capture the system's "state" in a compact drift-diffusion vector}

\item{The proposed model computes the drift and diffusion coefficients using closed-form algebraic solutions (means and variances) rather than iterative optimization.This enables execution at frequencies of 1 Hz+, providing instantaneous risk quantification where iterative models (like VARIMA's grid search) would introduce unacceptable lag.}

\item{The proposed model does not require a large training corpus or cloud-based weights, it is fully functional as a self-contained unit onboard a vessel.}

\end{itemize}

\bibliographystyle{IEEEtran}
\bibliography{template}

@InProceedings{alexiou2005development,
	author    = {Alexiou, A. and Mathioudakis, K.},
	title     = {Development of Gas Turbine Performance Models Using a Generic Simulation Tool},
	booktitle = {Volume 1: Turbo Expo 2005},
	year      = {2005},
	pages     = {185--194},
	address   = {Reno, Nevada, USA},
	month     = jan,
	publisher = {ASMEDC}
}

@InProceedings{asgari2021recurrent,
	author    = {Asgari, Hamid and Ory, Emmanuel and Lappalainen, Jari},
	title     = {Recurrent Neural Network Based Simulation of a Single Shaft Gas Turbine},
	year      = {2021},
	pages     = {99--106},
	month     = mar,
	note      = {Proceedings pages 99--106}
}

@InProceedings{batayev2018prediction,
	author = {Batayev, Nurlan and Onbayev, Ayan},
	title  = {Prediction of gas turbine parameters based on machine learning regression methods},
	year   = {2018},
	pages  = {217--221},
	month  = jul
}

@Article{belov2020hybrid,
	author   = {Belov, Sergei and Nikolaev, Sergei and Uzhinsky, Ighor},
	title    = {Hybrid Data-Driven and Physics-Based Modeling for Gas Turbine Prescriptive Analytics},
	journal  = {International Journal of Turbomachinery, Propulsion and Power},
	year     = {2020},
	volume   = {5},
	number   = {4},
	pages    = {29},
	month    = dec,
	publisher = {Multidisciplinary Digital Publishing Institute}
}

@Article{dai2024novel,
	author   = {Dai, Shun and Zhang, Xiaoyi and Luo, Mingyu},
	title    = {A Novel Data-Driven Approach for Predicting the Performance Degradation of a Gas Turbine},
	journal  = {Energies},
	year     = {2024},
	volume   = {17},
	number   = {4},
	pages    = {781},
	month    = jan,
	publisher = {Multidisciplinary Digital Publishing Institute}
}

@Article{gazzetta2017real,
	author  = {Gazzetta~Junior, Henrique and Bringhenti, Cleverson and Barbosa, Jo~{a}o Roberto and Tomita, Jesuino Takachi},
	title   = {Real-Time Gas Turbine Model for Performance Simulations},
	journal = {Journal of Aerospace Technology and Management},
	year    = {2017},
	volume  = {9},
	number  = {3},
	pages   = {346--356},
	month   = aug
}

@InProceedings{panov2010gasturbolib,
	author    = {Panov, V.},
	title     = {GasTurboLib: Simulink Library for Gas Turbine Engine Modelling},
	year      = {2010},
	pages     = {555--565},
	month     = feb,
	publisher = {American Society of Mechanical Engineers Digital Collection}
}

@book{saravanamuttoo2007gas,
	author    = {Saravanamuttoo, H.~I.~H. and Cohen, Henry and Rogers, G.~F.~C.},
	title     = {Gas turbine theory},
	publisher = {Prentice Hall},
	year      = {2007},
	edition   = {5},
	address   = {Harlow},
	oclc      = {476334045}
}

@article{singh2018modeling,
	author  = {Singh, Richa and Maity, Arnab and Nataraj, P.~S.~V.},
	title   = {Modeling, Simulation and Validation of Mini {SR-30} Gas Turbine Engine},
	journal = {IFAC-PapersOnLine},
	year    = {2018},
	volume  = {51},
	number  = {1},
	pages   = {554--559},
	month   = jan
}

@article{zhigarev2021development,
	author  = {Zhigarev, V.~A. and Minakov, A.~V. and Guzei, D.~V. and Pryazhnikov, M.~I. and Panteleev, V.~I.},
	title   = {Development of a mathematical model of diesel-generator units with a valve-inductor generator},
	journal = {Journal of Physics: Conference Series},
	year    = {2021},
	volume  = {2057},
	number  = {1},
	pages   = {012076},
	month   = oct,
	url     = {https://iopscience.iop.org/article/10.1088/1742-6596/2057/1/012076}
}

@inproceedings{mohammad2021hybrid,
	author = {Mohammad, A. and Rezaei, R. and Hayduk, C. and Delebinski, T.~O. and Shahpouri, S. and Shahbakhti, M.},
	title  = {Hybrid Physical and Machine Learning-Oriented Modeling Approach to Predict Emissions in a Diesel Compression Ignition Engine},
	year   = {2021},
	pages  = {2021--01--0496},
	month  = apr,
	url    = {https://www.sae.org/content/2021-01-0496/}
}

@phdthesis{singh2023hybrid,
	author = {Singh, S.~K.},
	title  = {Hybrid Physics-Infused Machine Learning Framework For Fault Diagnostics and Prognostics in Cyber-Physical System Of Diesel Engine},
	school = {Dept. of Automotive Engineering, Clemson University},
	year   = {2023},
	url    = {https://www.proquest.com/openview/e0412f5729f806451dc25df4e76da380/1?pq-origsite=gscholar\&cbl=18750\&diss=y}
}

@article{gkerekos2019machine,
	author  = {Gkerekos, C. and Lazakis, I. and Theotokatos, G.},
	title   = {Machine learning models for predicting ship main engine Fuel Oil Consumption: A comparative study},
	journal = {Ocean Engineering},
	year    = {2019},
	volume  = {188},
	pages   = {106282},
	month   = sep,
	url     = {https://linkinghub.elsevier.com/retrieve/pii/S0029801819304561}
}

@article{willard2023integrating,
	author  = {Willard, J. and Jia, X. and Xu, S. and Steinbach, M. and Kumar, V.},
	title   = {Integrating Scientific Knowledge with Machine Learning for Engineering and Environmental Systems},
	journal = {ACM Computing Surveys},
	year    = {2023},
	volume  = {55},
	number  = {4},
	pages   = {1--37},
	month   = apr,
	url     = {https://dl.acm.org/doi/10.1145/3514228}
}

@book{kloeden2012numerical,
  title={Numerical solution of SDE through computer experiments},
  author={Kloeden, Peter Eris and Platen, Eckhard and Schurz, Henri},
  year={2012},
  publisher={Springer Science \& Business Media}
}

\end{document}